\documentclass[conference]{IEEEtran}
\IEEEoverridecommandlockouts
\usepackage{cite}
\usepackage{amsmath,amssymb,amsfonts}
\usepackage{algorithmic}
\usepackage{graphicx}
\usepackage{textcomp}
\usepackage{xcolor}
\usepackage{multirow}
\usepackage{multirow}     
\usepackage{array}    
\usepackage{url} 
\usepackage{colortbl}     
\usepackage{xcolor}       
\usepackage{booktabs}   
\usepackage{algorithm}
\usepackage{algorithmic}

\def\BibTeX{{\rm B\kern-.05em{\sc i\kern-.025em b}\kern-.08em
    T\kern-.1667em\lower.7ex\hbox{E}\kern-.125emX}}
\begin{document}

\title{PolicySimEval: A Benchmark for Evaluating Policy Outcomes through Agent-Based Simulation}


\author{
    \IEEEauthorblockN{Jiaju Kang\textsuperscript{1,*}\thanks{*Corresponding author: kjj\_python@163.com}} \quad Puyu Han\textsuperscript{2}  \quad Tian Zhang\textsuperscript{5} \quad Luqi Gong\textsuperscript{3,4}
    \IEEEauthorblockA{\textsuperscript{1} Beijing Normal University \hspace{2mm} \textsuperscript{2} Southern University of Science and Technology} \hspace{2mm} \textsuperscript{3} Zhejiang Lab \hspace{2mm} \textsuperscript{4} Beijing University of Posts and Telecommunications \hspace{2mm}
    \textsuperscript{5} ESIGELEC\\
}

\maketitle

\begin{abstract}
With the growing adoption of agent-based models in policy evaluation, a pressing question arises: Can such systems effectively simulate and analyze complex social scenarios to inform policy decisions? Addressing this challenge could significantly enhance the policy-making process, offering researchers and practitioners a systematic way to validate, explore, and refine policy outcomes. To advance this goal, we introduce PolicySimEval, the first benchmark designed to evaluate the capability of agent-based simulations in policy assessment tasks. PolicySimEval aims to reflect the real-world complexities faced by social scientists and policymakers. The benchmark is composed of three categories of evaluation tasks: (1) 20 comprehensive scenarios that replicate end-to-end policy modeling challenges, complete with annotated expert solutions; (2) 65 targeted sub-tasks that address specific aspects of agent-based simulation (e.g., agent behavior calibration); and (3) 200 auto-generated tasks to enable large-scale evaluation and method development. Experiments show that current state-of-the-art frameworks struggle to tackle these tasks effectively, with the highest-performing system achieving only 24.5\% coverage rate on comprehensive scenarios, 15.04\% on sub-tasks, and 14.5\% on auto-generated tasks. These results highlight the difficulty of the task and the gap between current capabilities and the requirements for real-world policy evaluation.
\end{abstract}

\begin{IEEEkeywords}
Agent-Based Modeling, Policy Simulation Benchmark, Reproducibility and Evaluation, Complex Social Systems
\end{IEEEkeywords}

\section{Introduction}
\label{sec:intro}

Policy evaluation is essential for societal decision-making, requiring advanced tools to simulate and analyze complex policy scenarios. Agent-based models (ABMs) have emerged as a promising approach, capturing individual behaviors, group dynamics, and emergent phenomena beyond traditional deterministic models.\cite{Kremmydas2012AgentBM} However, their effectiveness in informing policy decisions remains uncertain, posing both academic and practical challenges.\cite{doi:10.1073/pnas.072079399}

ABMs for policy evaluation face significant real-world application issues, such as calibrating agent behaviors, integrating heterogeneous datasets, and interpreting results.\cite{hammond2015considerations} These tasks demand sophisticated technical capabilities and domain expertise. Additionally, the lack of standardized benchmarks hinders the systematic assessment of these models' performance and limitations.\cite{blume2015agent}

To address these challenges, we introduce \textbf{PolicySimEval}, the first benchmark specifically designed to evaluate ABMs in policy assessment tasks. \textbf{PolicySimEval} aims to:

\begin{itemize}
  \item Provide a rigorous framework to assess ABMs on real-world policy challenges.
  \item Facilitate the development and refinement of agent-based simulation methods by identifying their strengths and weaknesses.
\end{itemize}

\textbf{PolicySimEval} comprises three categories of evaluation tasks:

\begin{itemize}
  \item \textbf{Comprehensive Scenarios (20)}: End-to-end tasks replicating real-world policy challenges, accompanied by expert-annotated solutions.
  \item \textbf{Targeted Sub-tasks (65)}: Specific challenges such as agent behavior calibration, data integration, and hyperparameter tuning to test detailed aspects of ABMs.
  \item \textbf{Auto-Generated Tasks (200)}: Diverse, programmatically created problems to enable large-scale development, fine-tuning, and experimentation.
\end{itemize}

To ensure robust evaluation, \textbf{PolicySimEval} incorporates a variety of metrics that balance quantitative indicators (e.g., outcome alignment with objectives) and qualitative measures (e.g., interpretability of agent interactions and adaptability to unforeseen challenges). This multi-dimensional approach provides a comprehensive understanding of the models’ capabilities and limitations.

Our experiments demonstrate that current state-of-the-art frameworks struggle with \textbf{PolicySimEval}. The best-performing system achieves only 24.5\% accuracy on comprehensive scenarios and 15.04\% on targeted sub-tasks, highlighting the gap between existing tools and the requirements for effective policy evaluation. These findings underscore the need for further innovation in agent-based simulation systems.\cite{ZHAO20112189}

In summary, \textbf{PolicySimEval} serves as a foundational benchmark to advance research in agent-based simulation for policy evaluation. By offering a structured and challenging framework, it aims to drive the development of tools and methodologies that enhance policy decision-making and deepen the understanding of complex social systems.

\begin{figure*}[t]
        \centering
	\includegraphics[width=1.9\columnwidth]{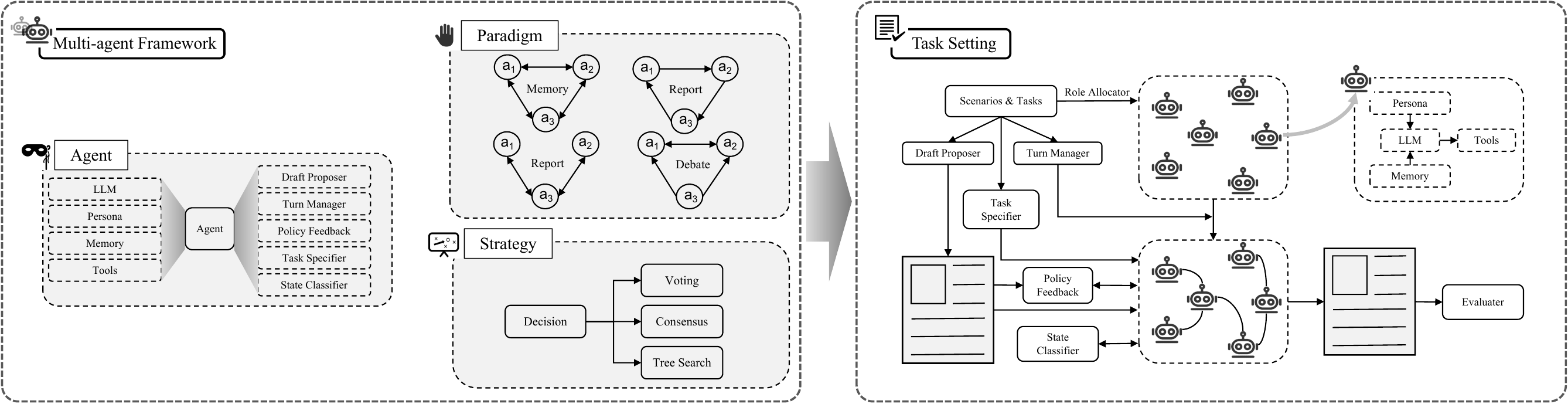} 
	\caption{Inspired by \cite{becker2024multiagentlargelanguagemodels}, we adopted this open-source and efficient multi-agent framework as the foundation for multi-agent communication in task construction.}
	\label{fig_ecg_ins}
\end{figure*}

\section{Related Work}

\subsection{Policy Simulation Frameworks and Benchmarks}

Policy simulation is a key area in computational social science, traditionally utilizing deterministic models like system dynamics that simulate macro-level phenomena but overlook micro-level interactions. Recent advancements have introduced agent-based frameworks such as MASON \cite{MASON}, Repast \cite{Collier2001}, and NetLogo \cite{tisue2004netlogo}, which enable the simulation of individual behaviors and their aggregate effects in dynamic social environments. However, these tools primarily offer flexible modeling environments without evaluating simulation accuracy, reproducibility, or generalizability. Existing benchmarks lack unified standards; platforms like CoMSES Net facilitate sharing ABMs but do not assess their correctness or reproducibility in real-world policy scenarios\cite{Windrum2007EmpiricalVO}. Similarly, frameworks like OpenABM emphasize accessibility and collaboration without establishing systematic evaluation protocols.\cite{Hinch2021} These limitations highlight the need for a dedicated benchmark to assess agent-based simulations in realistic and diverse policy challenges\cite{janssen2006empirically}.

\subsection{Agent-Based Modeling and Evaluations}

Agent-based modeling (ABM) is a prominent approach for studying complex systems, enabling the exploration of individual decision-making, heterogeneous interactions, and emergent phenomena. ABMs are particularly valuable in policy domains such as urban planning, epidemic control, and economic policy interventions\cite{GONZALEZMENDEZ2021105110,An2017,AN2021109685}. However, evaluating ABMs remains challenging, as traditional methods like manual expert validation or comparison to real-world data are labor-intensive and lack reproducibility. Automated evaluation methods, including Bayesian calibration\cite{thober2017, Lamperti2018} and ensemble modeling \cite{https://doi.org/10.1111/ajae.12174}, improve reliability but require significant computational resources and domain-specific expertise\cite{HUBER2018143}. Integrating machine learning with ABMs, such as using reinforcement learning to train agents or employing generative models to synthesize agent behaviors, has enhanced their performance and usability. Despite these advancements, comprehensive benchmarks for evaluating ABM-based policy tasks, particularly those addressing the end-to-end process of configuring, running, and interpreting simulations in diverse scenarios, are largely absent.

\section{PolicySimEval}

\begin{table*}[ht]
\centering
\scriptsize 
\renewcommand{\arraystretch}{1.6} 
\setlength{\tabcolsep}{10pt} 
\caption{Evaluation Metrics and Their Calculation Methods}
\begin{tabular}{p{3.5cm}|p{3.5cm}|p{8cm}}
\hline
\rowcolor[HTML]{EFEFEF} 
\textbf{Category}       & \textbf{Metric}           & \textbf{Formula and Description}                                                                                     \\ \hline

\multirow{6}{3.5cm}{\textbf{Argument Coverage}} 
& \multirow{2}{3.5cm}{Task Completion}             
& $C_t = \frac{S_c}{S_t}$ \\ 
&  
& Measures whether the system completes all required steps. $S_c$ is the number of completed steps, and $S_t$ is the total steps. \\ \cline{2-3} 

& \multirow{2}{3.5cm}{Discussion Completeness}     
& $D_c = \frac{|A_s \cap A_g|}{|A_g|}$ \\ 
&  
& Evaluates the overlap between system-generated arguments ($A_s$) and gold-standard arguments ($A_g$). \\ \cline{2-3} 

& \multirow{2}{3.5cm}{Coverage Rate}              
& $R_{cover} = \text{BLEU}(\text{Generated\_Text}, \text{Reference\_Text})$ \\ 
&  
& Measures the similarity between generated and reference texts using BLEU or ROUGE scores. \\ \hline

\multirow{6}{3.5cm}{\textbf{Behavior Calibration}} 
& \multirow{2}{3.5cm}{Behavior Consistency}   
& $E_b = \frac{1}{N} \sum_{i=1}^N ||B_i - \hat{B_i}||$ \\ 
&  
& Quantifies the error in calibrating agent behaviors, where $B_i$ is the true behavior rule, and $\hat{B_i}$ is the generated rule. \\ \cline{2-3} 

& \multirow{2}{3.5cm}{Group Coordination}    
& $H_c = -\sum_{k=1}^K P_k \log P_k$ \\ 
&  
& Evaluates group behavior coordination using entropy, where $P_k$ is the probability of the $k$-th behavior type, and $K$ is the total behavior types. \\ \cline{2-3} 

& \multirow{2}{3.5cm}{Trajectory Interpretability} 
& $T_{sim} = \frac{1}{M} \sum_{j=1}^M \text{Cosine\_Similarity}(T_{j,s}, T_{j,g})$ \\ 
&  
& Measures similarity between generated ($T_{j,s}$) and gold-standard ($T_{j,g}$) trajectories. $M$ is the number of trajectory samples. \\ \hline

\multirow{6}{3.5cm}{\textbf{Language and Ethics}} 
& \multirow{2}{3.5cm}{Language Quality}            
& $Q_l = \frac{1}{N} \sum_{i=1}^N (\text{Grammar\_Score}(T_i) + \text{Semantic\_Score}(T_i))$ \\ 
&  
& Assesses the grammar and semantic quality of text $T_i$. \\ \cline{2-3} 

& \multirow{2}{3.5cm}{Ethical Risk}     
& $R_e = \frac{N_b}{N_t}$ \\ 
&  
& Measures the proportion of texts containing biased expressions, combining lexicon-based matching and semantic detection via large language models. \\ \cline{2-3} 

& \multirow{2}{3.5cm}{Ethical Compliance}          
& $C_e = \text{ROUGE}(\text{Generated\_Text}, \text{Ethical\_Standard\_Text})$ \\ 
&  
& Evaluates the similarity between generated texts and ethical standards. \\ \hline

\multirow{4}{3.5cm}{\textbf{Outcome Effectiveness}} 
& \multirow{2}{3.5cm}{Outcome Alignment}           
& $A_r = 1 - \frac{\sum_{i=1}^M ||O_i - \hat{O_i}||}{M}$ \\ 
&  
& Measures the deviation between generated outcomes ($\hat{O_i}$) and ideal outcomes ($O_i$) over $M$ tasks. \\ \cline{2-3} 

& \multirow{2}{3.5cm}{Dynamic Adjustment} 
& $T_a = \text{Time}(\text{Updated\_Result}) - \text{Time}(\text{Input\_Change})$ \\ 
&  
& Calculates the time taken to update results after input changes. \\ \hline

\multirow{4}{3.5cm}{\textbf{System Performance}} 
& \multirow{2}{3.5cm}{Response Time}               
& $T_r = \frac{\sum_{j=1}^N T_j}{N}$ \\ 
&  
& Evaluates average task completion time, where $T_j$ is the time for task $j$, and $N$ is the total tasks. \\ \cline{2-3} 

& \multirow{2}{3.5cm}{Stability Index}             
& $S_v = \max(F) - \min(F)$ \\ 
&  
& Measures the range of performance fluctuations ($F$), such as accuracy or response time, over a given period. \\ \hline

\end{tabular}
\label{tab:eval_metrics}
\end{table*}

\textbf{PolicySimEval} comprises three core components designed to address various dimensions and complexities of policy evaluation tasks. The first component, Comprehensive Scenarios, includes 20 end-to-end policy modeling tasks that simulate real-world evaluation processes, covering the entire workflow from policy goal setting to outcome analysis. These expert-calibrated scenarios are accompanied by detailed reference solutions to ensure scientific rigor and reliability\cite{Lee2015complexities}. The second component, Targeted Sub-tasks, consists of 65 tasks that focus on specific challenges in agent-based modeling, such as agent behavior calibration, multimodal data integration, and hyperparameter optimization. These fine-grained evaluations identify technical bottlenecks and precisely measure the quality of agent-based systems\cite{Robinson2012}. The third component, Auto-generated Tasks, features 200 programmatically created scenarios that offer diverse scales and policy contexts, supporting large-scale experimentation, method development, model fine-tuning, and comparative analysis\cite{Shen2024}.

\subsection{Data Augmentation Pipeline}

The data generation process is central to the design of the PolicySimEval benchmark, ensuring that scenarios are grounded in real-world complexity. Data is sourced from reliable repositories such as public policy documents, historical case studies, academic research, and statistical databases\cite{Zhang2019}. These sources are filtered to retain only high-quality information with rich contextual relevance, covering diverse domains like economics, public health, and environmental policy\cite{Guo2022}. The data is then cleaned to remove noise and redundancies, followed by structuring into formats such as tables, knowledge graphs, or textual summaries to facilitate computational use.

To enrich scenario diversity, data augmentation techniques are applied, including the expansion of contextual information and the creation of synthetic and adversarial cases.\cite{Fagiolo2016} These cases are designed to evaluate system performance under varying conditions, including extreme scenarios that test robustness. Special attention is given to risk assessment data to simulate potential challenges prior to policy implementation.

Finally, gold-standard answers are manually curated by domain experts to ensure accuracy and consistency.\cite{COCKRELL2017157} These standards are periodically updated to reflect changes in policy contexts, maintaining the benchmark's relevance for evaluating policy simulation methods.

\subsection{Evaluation Metrics}
To comprehensively evaluate the performance of agent-based systems in policy simulation, we propose a structured framework of evaluation metrics. These metrics are organized into five categories: \textit{Argument Coverage}, \textit{Behavior Calibration}, \textit{Language and Ethics}, \textit{Outcome Effectiveness}, and \textit{System Performance}. Each category addresses a specific aspect of system capabilities, such as task completion, ethical compliance, and system efficiency\cite{Murić2022}. 

The metrics are carefully designed to balance mathematical rigor with practical interpretability. For instance, argument coverage focuses on task completion rates and argument overlap with gold standards, while behavior calibration evaluates individual and group dynamics through entropy and trajectory similarity. Language and ethical risks are quantified using both lexical and semantic models, ensuring a thorough analysis of text quality and bias\cite{Chen02102017}.

Table~\ref{tab:eval_metrics} summarizes the metrics, including their mathematical formulations and descriptive interpretations. This tabular representation not only facilitates clarity but also serves as a quick reference for researchers implementing these metrics in their systems.

\section{Experiment}

This experimental design aims to evaluate the performance of different agent models in policy evaluation tasks, following strict control conditions and multidimensional evaluation criteria. The execution time for each task is limited to 40 minutes (excluding API response time), with a token limit applied. The token limit for specific sub-tasks is set to 400k, while for comprehensive scenario tasks and automated generation tasks, the token limit is 600k.

The experimental tasks are categorized into three types: comprehensive scenario tasks, specific sub-tasks, and automated generation tasks. \textbf{Task 1: Comprehensive scenario tasks} consist of 20 end-to-end simulated tasks that replicate real-world policy challenges and are accompanied by expert solutions. These tasks aim to evaluate the agent's overall capability in complex policy contexts. \textbf{Task 2: Specific sub-tasks,} totaling 65 tasks, focus on testing the agent's performance in specific technical aspects, such as agent behavior calibration, data integration, and hyperparameter tuning. \textbf{Task 3: Automated generation tasks}, consisting of 200 large-scale tasks, are designed to evaluate and advance the automation of methods, enhance task execution scalability, and improve the adaptability and generalization capabilities of the agent.

We experiment with two foundational large language models: the commercial model GPT-4o (gpt-4o-2024-08-06) and the open-source Llama 3.1 70B model, developed by Meta and served through https://www.together.ai/. These large language models provide powerful reasoning and text generation capabilities to support the agent in performing various types of policy evaluation tasks.

In terms of agent construction, we experimented with ReAct\cite{yao2023reactsynergizingreasoningacting} as a baseline agent. The ReAct agent iteratively prompts the underlying large language model (LLM) to output both an action and a natural language "thought," using the interaction history as context. At each step, the generated action is executed in the environment, and a <thought, action, observation> tuple is added to the history, until the agent submits an answer or exceeds token or compute limitations. We also explored ReAct-RAG (ReAct with Retrieval-Augmented Generation). Unlike the traditional ReAct agent, ReAct-RAG introduces a web retrieval component, enabling the agent to fetch additional information from the internet in real-time, thus enhancing its reasoning and knowledge expansion capabilities. Specifically, at each decision point, ReAct-RAG generates an action based on historical interactions and current context, while also using an external retrieval mechanism to search for relevant knowledge from the internet, which is then provided as input to the LLM.

To comprehensively assess the performance of the agent models, we adopt a set of evaluation metrics covering five main categories: Argument Coverage, Behavior Calibration, Language and Ethics, Outcome Effectiveness, and System Performance. The Argument Coverage category primarily evaluates whether the agent completes the required steps, the completeness of generated arguments, and the similarity between the generated text and reference text. Behavior Calibration measures the consistency of agent behaviors, group coordination, and the similarity between generated and gold-standard trajectories. Language and Ethics assesses the grammatical and semantic quality of the generated text, ethical risks, and compliance with ethical standards. Outcome Effectiveness evaluates the deviation between generated and ideal outcomes, as well as the time required to update results after input changes. Finally, System Performance focuses on evaluating the response time and stability of the system. These comprehensive evaluation metrics provide a multidimensional framework for examining and analyzing the agent models' task execution capabilities.
\section{Results}

\begin{table}[htbp]
\centering
\caption{Results of our baselines on Argument Coverage.}
\begin{tabular}{c|cc|ccc}
\toprule
\textbf{Task} & \textbf{Agent} & \textbf{Model} & \textbf{$C_t$} & \textbf{$D_c$} & \textbf{$R_{cover}$} \\
\midrule
\multirow{4}{*}{Task 1} & \multirow{2}{*}{ReAct} & GPT-4o & 21.08 & 15.38 & 23.73 \\ 
 &  & Llama 3.1 70B & 11.77 & 13.76 & 2.33 \\
 & \multirow{2}{*}{ReAct-RAG} & GPT-4o & 23.73 & 22.36 & 24.50 \\ 
 &  & Llama 3.1 70B & 13.11 & 19.65 & 16.28 \\
\midrule
\multirow{4}{*}{Task 2} & \multirow{2}{*}{ReAct} & GPT-4o & 15.53 & 3.99 & 10.07 \\ 
 &  & Llama 3.1 70B & 13.76 & 11.50 & 9.04 \\ 
 & \multirow{2}{*}{ReAct-RAG} & GPT-4o & 17.41 & 22.36 & 15.04 \\ 
 &  & Llama 3.1 70B & 14.20 & 19.65 & 12.88 \\ 
\midrule
\multirow{4}{*}{Task 3} & \multirow{2}{*}{ReAct} & GPT-4o & 8.06 & 17.41 & 13.42 \\
 &  & Llama 3.1 70B & 7.50 & 10.06 & 8.97 \\ 
 & \multirow{2}{*}{ReAct-RAG} & GPT-4o & 10.50 & 11.14 & 14.50 \\ 
 &  & Llama 3.1 70B & 9.50 & 7.75 & 9.32 \\ 
\bottomrule

\end{tabular}
\label{tab:1}
\end{table}

As shown in the table~\ref{tab:1}, the ReAct-RAG model, particularly with GPT-4o, outperforms others in task completion ($C_t$), discussion completeness ($D_c$), and coverage rate ($R_{cover}$). For example, in Task 1, GPT-4o achieves the highest $C_t$ and $D_c$, indicating better task fulfillment and more complete discussions compared to ReAct and Llama 3.1 70B. Similarly, it shows superior coverage with an $R_{cover}$ of 24.50 in Task 1.

\begin{table}[htbp]
\centering
\caption{Results of our baselines on Behavior Calibration.}
\begin{tabular}{c|cc|ccc}
\toprule
\textbf{Task} & \textbf{Agent} & \textbf{Model} & \textbf{$E_b$} & \textbf{$H_c$} & \textbf{$T_{sim}$} \\
\midrule
\multirow{4}{*}{Task 1} & \multirow{2}{*}{ReAct} & GPT-4o & 7.85 & 5.24 & 17.07 \\
 &  & Llama 3.1 70B & 7.85 & 0.69 & 5.78 \\
 & \multirow{2}{*}{ReAct-RAG} & GPT-4o & 19.24 & 18.29 & 19.32 \\
 &  & Llama 3.1 70B & 19.69 & 21.80 & 20.39 \\
\midrule
\multirow{4}{*}{Task 2} & \multirow{2}{*}{ReAct} & GPT-4o & 3.81 & 11.77 & 11.02 \\
 &  & Llama 3.1 70B & 2.80 & 7.37 & 13.95 \\
 & \multirow{2}{*}{ReAct-RAG} & GPT-4o & 3.94 & 15.71 & 16.90 \\
 &  & Llama 3.1 70B & 13.06 & 7.62 & 15.78 \\
\midrule
\multirow{4}{*}{Task 3} & \multirow{2}{*}{ReAct} & GPT-4o & 13.15 & 15.02 & 8.55 \\
 &  & Llama 3.1 70B & 5.28 & 8.10 & 3.15 \\
 & \multirow{2}{*}{ReAct-RAG} & GPT-4o & 17.39 & 20.00 & 15.24 \\
 &  & Llama 3.1 70B & 15.99 & 20.37 & 10.74 \\
\bottomrule

\end{tabular}
\label{tab:2}
\end{table}

As shown in the table~\ref{tab:2}, the ReAct-RAG model, particularly with GPT-4o, demonstrates superior performance in behavior calibration metrics ($E_b$, $H_c$, and $T_{sim}$). In Task 1, GPT-4o achieves the highest $E_b$ (19.24) and $T_{sim}$ (19.32), indicating better behavior calibration and trajectory similarity compared to ReAct and Llama 3.1 70B. In terms of group coordination, ReAct-RAG also performs well, with $H_c$ values of 18.29 for GPT-4o and 21.80 for Llama 3.1 70B, suggesting better coordination across tasks. The trend continues in Tasks 2 and 3, where ReAct-RAG outperforms other models, particularly in behavior consistency and trajectory similarity.

\begin{table}[htbp]
\centering
\caption{Results of our baselines on Language and Ethics.}
\begin{tabular}{c|cc|ccc}
\toprule
\textbf{Task} & \textbf{Agent} & \textbf{Model} & \textbf{$Q_l$} & \textbf{$R_e$} & \textbf{$C_e$} \\
\midrule
\multirow{4}{*}{Task 1} & \multirow{2}{*}{ReAct} & GPT-4o & 9.05 & 1.89 & 6.46 \\
 &  & Llama 3.1 70B & 15.74 & 1.75 & 0.60 \\
 & \multirow{2}{*}{ReAct-RAG} & GPT-4o & 19.07 & 4.11 & 9.24 \\
 &  & Llama 3.1 70B & 17.83 & 7.64 & 22.36 \\
\midrule
\multirow{4}{*}{Task 2} & \multirow{2}{*}{ReAct} & GPT-4o & 3.68 & 6.27 & 5.04 \\
 &  & Llama 3.1 70B & 9.10 & 2.70 & 12.46 \\
 & \multirow{2}{*}{ReAct-RAG} & GPT-4o & 22.54 & 6.44 & 15.78 \\
 &  & Llama 3.1 70B & 15.25 & 4.38 & 12.46 \\
\midrule
\multirow{4}{*}{Task 3} & \multirow{2}{*}{ReAct} & GPT-4o & 3.07 & 2.00 & 8.43 \\
 &  & Llama 3.1 70B & 9.24 & 3.00 & 8.78 \\
 & \multirow{2}{*}{ReAct-RAG} & GPT-4o & 18.25 & 4.31 & 16.78 \\
 &  & Llama 3.1 70B & 10.52 & 6.80 & 9.32 \\
\bottomrule
\end{tabular}
\label{tab:3}
\end{table}

As shown in the table~\ref{tab:3}, the ReAct-RAG model, especially with GPT-4o, outperforms others in language quality ($Q_l$) and ethical compliance ($C_e$). In Task 1, ReAct-RAG achieves the highest $Q_l$ (19.07) and $C_e$ (9.24), showing better language generation and ethical alignment compared to ReAct and Llama 3.1 70B. The trend continues in Tasks 2 and 3, where ReAct-RAG consistently leads in these metrics, particularly with GPT-4o.

\begin{table}[htbp]
\centering
\caption{Results of our baselines on Outcome Alignment and Dynamic Adjustment.}
\begin{tabular}{c|cc|cc}
\toprule
\textbf{Task} & \textbf{Agent} & \textbf{Model} & \textbf{$A_r$} & \textbf{$T_a$} \\
\midrule
\multirow{4}{*}{Task 1} & \multirow{2}{*}{ReAct} & GPT-4o & 0.76 & 30.00 \\
 &  & Llama 3.1 70B & 4.71 & 1.63 \\
 & \multirow{2}{*}{ReAct-RAG} & GPT-4o & 9.98 & 38.90 \\
 &  & Llama 3.1 70B & 15.26 & 33.46 \\
\midrule
\multirow{4}{*}{Task 2} & \multirow{2}{*}{ReAct} & GPT-4o & 3.45 & 20.57 \\
 &  & Llama 3.1 70B & 3.09 & 34.21 \\
 & \multirow{2}{*}{ReAct-RAG} & GPT-4o & 6.58 & 21.53 \\
 &  & Llama 3.1 70B & 7.49 & 45.31 \\
\midrule
\multirow{4}{*}{Task 3} & \multirow{2}{*}{ReAct} & GPT-4o & 15.71 & 7.87 \\
 &  & Llama 3.1 70B & 8.53 & 17.28 \\
 & \multirow{2}{*}{ReAct-RAG} & GPT-4o & 15.72 & 31.00 \\
 &  & Llama 3.1 70B & 11.19 & 22.12 \\
\bottomrule

\end{tabular}
\label{tab:4}
\end{table}

As shown in the table~\ref{tab:4}, the ReAct-RAG model outperforms the others in outcome alignment ($A_r$) and dynamic adjustment ($T_a$). In Task 1, ReAct-RAG with GPT-4o achieves the highest $A_r$ (9.98) and $T_a$ (38.90), indicating better alignment with ideal outcomes and faster updates compared to ReAct and Llama 3.1 70B. This trend continues in Tasks 2 and 3, where ReAct-RAG consistently leads in both metrics, especially with Llama 3.1 70B showing significant improvement in $T_a$.

\begin{table}[htbp]
\centering
\caption{Results of our baselines on System Performance.}
\begin{tabular}{c|cc|cc}
\toprule
\textbf{Task} & \textbf{Agent} & \textbf{Model} & \textbf{$T_r$} & \textbf{$S_v$} \\
\midrule
\multirow{4}{*}{Task 1} & \multirow{2}{*}{ReAct} & GPT-4o & 9.57 & 3.15 \\
 &  & Llama 3.1 70B & 12.03 & 6.02 \\
 & \multirow{2}{*}{ReAct-RAG} & GPT-4o & 29.86 & 16.96 \\
 &  & Llama 3.1 70B & 21.02 & 17.49 \\
\midrule
\multirow{4}{*}{Task 2} & \multirow{2}{*}{ReAct} & GPT-4o & 17.89 & 11.84 \\
 &  & Llama 3.1 70B & 8.64 & 2.36 \\
 & \multirow{2}{*}{ReAct-RAG} & GPT-4o & 24.12 & 16.59 \\
 &  & Llama 3.1 70B & 31.06 & 4.83 \\
\midrule
\multirow{4}{*}{Task 3} & \multirow{2}{*}{ReAct} & GPT-4o & 19.70 & 9.64 \\
 &  & Llama 3.1 70B & 34.87 & 3.59 \\
 & \multirow{2}{*}{ReAct-RAG} & GPT-4o & 30.96 & 9.97 \\
 &  & Llama 3.1 70B & 40.27 & 9.18 \\
\bottomrule
\end{tabular}
\label{tab:5}
\end{table}

As shown in the table~\ref{tab:5}, the ReAct-RAG model demonstrates superior system performance compared to the others, particularly with Llama 3.1 70B. In Task 1, ReAct-RAG with GPT-4o achieves the highest $T_r$ (29.86), indicating faster response times, while the $S_v$ (stability index) is also high (16.96), reflecting greater stability. In Task 2, ReAct-RAG with Llama 3.1 70B shows a significant improvement in $T_r$ (31.06) and a moderate $S_v$ (4.83), highlighting better overall system performance. Similarly, in Task 3, ReAct-RAG consistently outperforms the other models in both $T_r$ and $S_v$, with Llama 3.1 70B achieving the highest $T_r$ (40.27).

\section{Conclusion}

In this paper, we introduced PolicySimEval, the first benchmark designed to evaluate agent-based models (ABMs) in policy assessment tasks. By offering a structured and comprehensive framework, PolicySimEval aims to address the challenges faced by policymakers and researchers when using ABMs for complex social scenario simulations. We outlined three categories of evaluation tasks—comprehensive scenarios, targeted sub-tasks, and auto-generated tasks—each targeting different aspects of ABM performance. These tasks are designed to assess both the effectiveness of agent-based simulations and their ability to adapt to real-world policy contexts.

Our experiments with current state-of-the-art frameworks demonstrate that even the best-performing systems struggle to meet the requirements of PolicySimEval. The results indicate that the existing ABM tools face significant limitations, particularly in handling complex, multi-faceted policy challenges. The gap in performance highlights the urgent need for further innovation in simulation methodologies and suggests that agent-based modeling in policy contexts still has considerable room for improvement. This reinforces the importance of developing and refining tools like PolicySimEval that can guide progress in this field.

Looking forward, PolicySimEval provides a solid foundation for advancing research in ABM-based policy evaluation. Its rigorous structure and diverse evaluation metrics will continue to support the development of more effective simulation systems. Future work could involve expanding the benchmark to include additional policy domains, improving the scalability of evaluation tasks, and refining the metrics to better capture the nuances of agent behavior and decision-making processes. By fostering further innovation, PolicySimEval has the potential to play a crucial role in bridging the gap between ABM research and real-world policy applications.

\newpage

\bibliographystyle{IEEEbib}
\bibliography{icme2025references}

\vspace{12pt}
\color{red}

\end{document}